\documentclass[aps,prd,superscriptaddress]{revtex4}
\usepackage{graphicx}
\usepackage{amssymb}
\usepackage{hyperref}

\usepackage{ulem}

\newcommand{\sgn}{\mbox{sgn}\,}
\newcommand{\Si}{\mathop{\rm Si}\nolimits}

\def\(({\left(}
\def\)){\right)}
\def\[[{\left[}
\def\]]{\right]}
\def \<{\langle}
\def \>{\rangle}

\begin{document}

\title{Off-diagonal correlations of the Calogero-Sutherland model}

\author{G.E.~Astrakharchik}
\affiliation{Dipartimento di Fisica, Universit\`a di Trento and BEC-INFM, I-38050 Povo, Italy}
\affiliation{Institute of Spectroscopy, 142190 Troitsk, Moscow region, Russia}

\author{D.M.~Gangardt}
\affiliation{\mbox{Laboratoire de Physique Th\'eorique et Mod\`eles
Statistiques, Universit\'e Paris Sud, 91405 Orsay Cedex, France}}

\author{Yu.E.~Lozovik}
\affiliation{Institute of Spectroscopy, 142190 Troitsk, Moscow region, Russia}

\author{I.A.~Sorokin}
\affiliation{Institute of Spectroscopy, 142190 Troitsk, Moscow region, Russia}

\date{\today}

\begin{abstract}
We study correlation functions of the Calogero-Sutherland model in the whole range
of the interaction parameter. Using the replica method we obtain analytical
expressions for the long-distance asymptotics of the one-body density matrix in
addition to the previously derived asymptotics of the pair-distribution function
[D.M.~Gangardt and A.~Kamenev, Nucl.~Phys.~B, \textbf{610}, 578 (2001)]. The leading
analytic and non-analytic terms in the short-distance expansion of the one-body
density matrix are discussed. Numerical results for these correlation functions are
obtained using Monte Carlo techniques for all distances. The momentum distribution
and static structure factor are calculated. The potential and kinetic energies are
obtained using the Hellmann-Feynman theorem. Perfect agreement is found between the
analytical expressions and numerical data. These results allow for the description
of physical regimes of the Calogero-Sutherland model. The zero temperature phase
diagram is found to be of a crossover type and includes quasi-condensation,
quasi-crystallization and quasi-supersolid regimes.
\end{abstract}

\maketitle

\section{Introduction}

There is an ongoing interest in correlation properties of the Calogero-Sutherland
model (CSM). From the theoretical perspective the Calogero-Sutherland model provides
a rare example of exactly solvable model with relatively simple structure of
eigenstates. In the present work we address the question of coherence properties of
the CSM measured by the off-diagonal correlation functions. Our study complements
the previous results on the diagonal density-density correlation properties of the
CSM and alows to draw conclusions about the coexistence of both types of long-range
correlations in an appropriate interval of the interaction parameter.

Introduced in Ref.~\cite{Sutherland1971}, the CSM describes a system of particles
interacting with a scale-free potential, which is inversely proportional to the
square of the distance between particles. The ground state wave function was shown
to have a form which is factorizable over pairs of particles. Each factor is
proportional to $(x_i-x_j)^\lambda$ with $x_i, x_j$ being the coordinates of
particles forming a pair and the parameter $\lambda$ being directly related to the
interaction strength. In addition to the convenient description of the excitations
in terms of non-interacting particles with fractional statistics \cite{FracStat},
this particular form of the ground state suggested a possibility to study
correlations functions of this model.

Indeed, it was noted in \cite{Sutherland1971} that for three special values
$\lambda=1/2,1$ and $2$ the ground state probability of the CSM coincides with
the probability distribution of the eigenvalues of unitary random matrices, so
the early results of Dyson \cite{Dyson62} describe the static density
correlations in the CSM. The analogy with random matrix theory allows also the
calculation of dynamical density correlations \cite{Simons1993} and dynamical
Green's function \cite{Sutherland1992,HaldaneZirnbauer1993}. To deal with
other values of interactions the Jack polynomial method has been applied to
find correlation functions for integer \cite{Forrester1992} and rational
\cite{Ha1994} values of $\lambda$. The common drawback of these methods for a
rational $\lambda=p/q$ is that the final expression for the correlation
function is usually given as a sum over fractional excitations involving $p+q$
particle and hole quantum numbers, which goes over $p+q$ integrals in the
thermodynamic limit. Such a decomposition makes the result appear as a highly
irregular function of the coupling constant $\lambda$, leaving little hope of
approaching its irrational values.

Recently, there has been  success \cite{GangardtKamenev2001} in obtaining the
density-density correlation functions of the CSM for arbitrary coupling $\lambda$,
leading to transparent asymptotic expressions in the long-distance limit. Based on
the replica method from the theory of disordered systems, this approach involves the
representation of the correlation functions using a duality transformation as an
$m$-dimensional integral with eventual analytic continuation in $m$. Similar methods
were applied recently \cite{Gangardt2004} to study off-diagonal correlations of
one-dimensional impenetrable bosons, equivalent to the bosonic CSM for one specific
value of $\lambda=1$. For recent progress in calculation of correlation functions
of integrable models such as Heisenberg spin chain see
\cite{SpinChainCorr} and references therein. Results  for delta-interacting
bosons can be found in the book \cite{KorepinBook}.  

In this paper we extend the replica method for studying the off-diagonal
correlations (equal-time Green's function) and corresponding momentum distributions.
The main result is the exact long-distance asymptotic behaviour of the one-body density
matrix in the form of the Haldane's universal hydrodynamic expansion
\cite{Haldane1981}. We consider both bosonic and fermionic statistics of the
particles encoded in the symmetry of the wavefunctions using the definitions
of the original  work of Sutherland \cite{Sutherland1971} 
(see also \cite{Sutherland1992}). Being irrelevant for
the density correlations, the quantum statistics affects drastically the
results for the off-diagonal correlations.

In addition, we study the short-distance behaviour of the one-body density
matrix and find several leading terms (analytic and non-analytic) in the
short-distance expansion.  The latter is directly related to the high-momentum
tails of the momentum distribution. The short-distance (high momentum) physics
enters the expressions for potential and kinetic energies which we calculate
as a function of interactions by using the Hellmann-Feynman theorem.

To check our predictions we use a Monte Carlo method to calculate numerically the
correlation functions for arbitrary distances and different values of the
interaction parameter. The advantage of the Calogero-Sutherland model is that its
ground state wavefunction is known explicitly and can be easily sampled by
Metropolis algorithm. This permits us to obtain unambiguous results for intermediate
distances, where analytical methods fail.

Combining these results with previous knowledge of the diagonal two-body
correlations (pair distribution function) \cite{GangardtKamenev2001} allows us to
describe different physical regimes of the CSM at zero temperature. In particular,
we discuss long- and short-range order as a function of the coupling constant
$\lambda$ for diagonal and off-diagonal correlations. Since the true long-range
order is absent in one dimension, we define it through the correlation function of
the (local) order parameter which has the slowest power-law decay and use the word
``quasi'' to stress this peculiarity of one dimension. We propose a phase diagram
which describes (in order of increasing interaction strength) the crossover between
three physical regimes: the quasi-condensate, quasi-supersolid and quasi-crystal.

The paper is organized as follows. In Section~\ref{sec:CSM} we introduce the
Hamitltonian of the Calogero-Sutherland model, its solution for the ground state and
define correlation functions of interest. In Section~\ref{sec:obdm} we present the
calculation of the one-body density matrix based on the replica method and discuss
the thermodynamic limit in Section~\ref{sec:thermo_lim}. Results for the
short-distance behaviour of the one-body density matrix and discussion of the
kinetic and potential energies are found in Section~\ref{sec:shortcorr}.
Section~\ref{sec:num_res_disc} is devoted to numerical Monte Carlo simulations and
discussion of the physics of the CSM. In section~\ref{sec:concl} we propose the
phase diagram of the CSM and draw our conclusions. Appendix~\ref{app:analAn}
contains mathematical details of our calculations.

\section{The Calogero-Sutherland model\label{sec:CSM}}

We consider a finite system of $N$ particles of mass $m$ on a ring of length $L$.
The Hamiltonian of the Calogero-Sutherland model is given by the sum of
the kinetic energy
and pair interactions controlled by parameter $\lambda$:
\begin{eqnarray} \label{ham}
H=\frac{\hbar^2}{2m}\left[-\sum_{i=1}^N\frac{\partial^2}{\partial x_i^2}
+\lambda(\lambda-1)\sum_{i\neq j} \frac{\pi^2/L^2}{\sin^2(\pi(x_i-x_j)/L)}\right].
\end{eqnarray}
The interaction between two particles on the ring is inversely proportional to
square of the chord distance between them. It becomes  an inverse square potential in
the thermodynamic limit.

The ground state wave function of the Hamiltonian (\ref{ham}) was found in
\cite{Sutherland1971} and can be written as:
\begin{eqnarray}\label{gs}
\Phi (x_1,\ldots, x_N) = C_N (\lambda) \prod_{k<l} \left(e^{2\pi i
x_k/L}-e^{2\pi i x_l/L}\right)^\lambda.
\end{eqnarray}
Here $C_N(\lambda)$ is the normalization constant given by
\begin{eqnarray}
C^2_N (\lambda) = \frac{1}{L^N}\frac{\Gamma(1+\lambda)^N}{\Gamma(1+\lambda N)}.
\end{eqnarray}

The expression (\ref{gs}), as it stands, is valid only for a particular ordering of
particles, for instance $x_1<x_2<\ldots < x_N$. To extend the expression (\ref{gs})
to other ordering configuration one has to specify the quantum statistics
(bosonic
or fermionic) of the particles. We modify the expression (\ref{gs}) to take into
account the symmetry under permutation of the particle coordinates
\begin{eqnarray}\label{gs_mod}
\Phi_{B,F} (x_1,\ldots, x_N) = C_N (\lambda) \prod_{k<l}
\left|e^{2\pi i x_k/L}-e^{2\pi i x_l/L}\right|^\lambda S_{B,F},
\label{Phi0}
\end{eqnarray}
by introducing the factor $S_{B,F}$ such that for bosons $S_B=1$, while for fermions
$S_F=(-1)^P$ is the parity of the permutation $P$, where $x_{P_1}<x_{P_2}<\ldots
<x_{P_N}$. As we shall see different quantum statistics of particles is crucial for
off-diagonal correlations, which was already noted in the context of the
Calogero-Sutherland model in early works \cite{Sutherland1971}.

The main quantity of interest is the one-body density matrix which in terms of the
ground state wave function is written as
\begin{eqnarray} \label{g1_def}
g_1^{B,F}(x-y) = N \int_0^L d^{N-1} x \;\Phi^*_{B,F} (x_1,\ldots, x_{N-1},x)
                 \Phi_{B,F} (x_1,\ldots, x_{N-1},y) .
\end{eqnarray}
Due to the translational invariance in a homogeneous system the one-body density
matrix is a function of the difference $x-y$ only. Knowledge of the
one-body density
matrix enables one to calculate the momentum distribution as the Fourier transform:
\begin{eqnarray} \label{nk_def}
n_k = \int dx\; e^{-ikx} g_1 (x).
\label{nk}
\end{eqnarray}
The one-body density matrix has dimensions of density and is normalized so that
$g_1(0) = n$, where $n=N/L$ is the particle density. The momentum distribution $n_k$
is dimensionless, it is defined for $k_l=2\pi l/L$, where $l$ is an integer, and is
normalized to the total number of particles $\sum_k n_k = L g_1 (0) =N$.

In addition to the off-diagonal correlation functions we consider the two-body
density matrix (pair distribution function). It is defined as
\begin{eqnarray} \label{g2_def}
g_2(x-y) = N(N-1) \int d^{N-2}x\;\left|\Phi(x_1,\ldots, x_{N-2},x,y)\right|^2 .
\end{eqnarray}
The static structure factor is related to $g_2(x)$:
\begin{eqnarray} \label{sk_def}
S_k = 1+\frac{1}{n}\int dx\; e^{-ikx} (g_2(x)-n^2).
\end{eqnarray}
The two-body density matrix has dimensions of density squared and is normalized so
that $\lim_{x\to\infty} g_2(x) = n^2$, while the static structure factor is a
dimensionless quantity and $\lim_{k\to \infty} S_k = 1$. It is independent of
statistics, as it involves only the absolute value of the ground state wavefunction.
In the next section we calculate analytically the one-body density matrix
(\ref{g1_def}). The results for the pair distribution function (\ref{g2_def}) were
obtained in \cite{GangardtKamenev2001} and we reproduce them in the
Section~\ref{num:pair_dist}.

\section{One-body density matrix \label{sec:obdm}}

To calculate the one-body density matrix we define the dimensionless function
$G_1(\alpha)$ such that $g_1(x) = n G_1 (2\pi x/L)$ and $G_1(0)=1$. Factorizing the
ground state wave function (\ref{gs_mod}) we rewrite the definition (\ref{g1_def})
in the form of the average
\begin{eqnarray}
  \label{eq:G1}
  G_1(\alpha) = \frac{\Gamma(\lambda)\Gamma(1+\lambda N) }{2\pi
  \Gamma(\lambda(1+N))} \left\langle \prod_{j=1}^N |1-e^{i\theta_j}|^\lambda
  |e^{i\alpha}-e^{i\theta_j}|^\lambda\right\rangle_{N,\lambda},
\end{eqnarray}
where the average is defined as
\begin{eqnarray}
  \label{eq:average_def}
  \left\langle f\left(e^{i\theta_1},e^{i\theta_2},
      \ldots,e^{i\theta_N}\right)\right\rangle_{N,\lambda}= \frac{\Gamma^N
      (1+\lambda)}{\Gamma(1+\lambda N)}\int_0^{2\pi}
      \frac{d^N\theta}{(2\pi)^N} \left|\Delta_N
      (e^{i\theta})\right|^{2\lambda} f\left(e^{i\theta_1},e^{i\theta_2},
      \ldots,e^{i\theta_N}\right),
\end{eqnarray}
and $\Delta_N (z)$ is the Vandermonde determinant
\begin{eqnarray}
  \label{eq:vandermonde}
  \Delta_N (z) = \Delta(z_1,z_2,\ldots,z_N) =\prod_{i<j} (z_i-z_j).
\end{eqnarray}
Here we have changed the number of particles from $N$ to $N+1$ in order to deal with
$N$ dimensional integrals. This difference does not matter in the thermodynamic
limit, and for the finite system we will restore the correct number of particles in
the final expressions.

To calculate the average in (\ref{eq:G1}) we use the replica trick, along the lines
of the calculation in \cite{GangardtKamenev2001} and \cite{Gangardt2004}. Namely,
consider the following function
\begin{eqnarray}
  \label{eq:Z_def}
  Z^{(\lambda)}_m (\alpha) = \left\langle \prod_{j=1}^N (1-e^{i\theta_j})^m
  (e^{i\alpha}-e^{i\theta_j})^m\right\rangle_{N,\lambda}.
\end{eqnarray}

It can be shown along the lines of \cite{Kurchan1991} that $G_1(\alpha)$ is obtained
from $Z^{(\lambda)}_m (\alpha)$ by the analytical continuation $m\to\lambda$. Later we
discuss this procedure in some detail and show how the quantum statistics of the
particles, bosonic or fermionic, appears naturally in our calculations. For the
moment we take advantage of the duality transformation \cite{Duality}, which
enables one
to re-express the $N$-dimensional integral (\ref{eq:Z_def}) depending on the
parameter $m$ as a $m$-dimensional integral depending on $N$ as a parameter:
\begin{eqnarray}
  \label{eq:Z_dual}
  Z^{(\lambda)}_m (t) = \frac{e^{-iNm\alpha/2}}{S_m(1/\lambda)} \int_0^1 d^m x
  \; |\Delta_m(x)|^{\frac{2}{\lambda}} \prod_{a=1}^m x_a^{\frac{1}{\lambda}-1}
  (1-x_a)^{\frac{1}{\lambda}-1} (1-(1-e^{i\alpha})x_a)^N .
\end{eqnarray}
The duality $\lambda\leftrightarrow 1/\lambda$ becomes evident by comparing the power of
Vandermonde determinants in Eqs.~(\ref{eq:average_def}),~(\ref{eq:Z_def}) and
(\ref{eq:Z_dual}). We put $Z^{(\lambda)}_m (0)=1$, so the normalization
constant is given by the Selberg integral:
\begin{eqnarray}
  \label{eq:s_selberg}
  S_m(1/\lambda) = \int_0^1d^m x \; |\Delta_m(x)|^{\frac{2}{\lambda}}
\prod_{a=1}^m x_a^{\frac{1}{\lambda}-1} (1-x_a)^{\frac{1}{\lambda}-1} =
\prod_{a=1}^m \frac{\Gamma^2\left(\frac{a}{\lambda}\right)
\Gamma\left(1+\frac{a}{\lambda}\right)}{\Gamma\left(1+\frac{1}{\lambda}\right)
\Gamma\left(\frac{m+a}{\lambda}\right)}.
\end{eqnarray}

The dual representation is an excellent starting point for the asymptotic expansion
in the large $N$ limit. In this limit the main contribution to the integral over
each variable $x_a$ comes from the limits of integration $x_+=1$ and $x_-=0$, so we expand
close to these points:
\begin{eqnarray}
  x_a&=&x_-+\frac{\xi_a}{N(1-e^{i\alpha})}, \qquad \;\;\;\;a=1,\ldots,l \\
    x_b&=&x_+ -\frac{\xi_b}{N(1-e^{-i\alpha})}, \qquad b=l+1,\ldots,m .
\label{eq:saddle_points}
\end{eqnarray}

To the leading order in $N$ the stationary action is
\begin{eqnarray}
  \label{eq:saddle_action}
  (1-(1-e^{i\alpha}) x_a)^N = \left\{
  \begin{array}{ll}
    e^{-\xi_a}, & a=1,\ldots,l\\ e^{iN\alpha} e^{-\xi_a}, & a=l+1,\ldots,m
  \end{array}\right. ,
\end{eqnarray}
and the Vandermonde determinants factorize as
\begin{eqnarray}
  \label{eq:vandermonde_fact}
  \Delta_m (x) \simeq \left(\frac{1}{N(1-e^{i\alpha})}\right)^{l(l-1)/2}
  \left(\frac{1}{N(1-e^{-i\alpha})}\right)^{(m-l)(m-l-1)/2} \Delta_l (\xi_a)
  \Delta_{m-l} (\xi_b) ,
\end{eqnarray}
which allows the calculation of the remaining fluctuation contributions in the form of
the Selberg integrals \cite{MehtaRandMatr} as follows
\begin{eqnarray}
  \label{eq:i_lambda}
  I_l (1/\lambda) = \int_0^\infty d^l\xi\; |\Delta_l(\xi)|^\frac{2}{\lambda}
  \prod_{a=1}^l\xi_a^{\frac{1}{\lambda}-1} e^{-\xi_a} = \prod_{a=1}^l
  \frac{\Gamma\left(\frac{a}{\lambda}\right)\Gamma\left(1+\frac{a}{\lambda}\right)}
  {\Gamma\left(1+\frac{1}{\lambda}\right)}.
\end{eqnarray}

Collecting factors arising from the change of variables (\ref{eq:saddle_points}) and
summing over all $2^m$ saddle points yields the following result:
\begin{eqnarray}
  \label{eq:z_sumsaddle}
  Z^{(\lambda)}_m (\alpha) =\sum_{l=0}^m (-1)^{\frac{m}{\lambda}
  \left(\frac{m}{2}-l\right)} H^l_m (\lambda) \frac{
  e^{i\left(N+\frac{m}{\lambda}\right)\left(\frac{m}{2}-l\right)\alpha}}{
  \left(2N\sin\frac{\alpha}{2}\right)^\frac{l^2+(m-l)^2}{\lambda}}   ,
\end{eqnarray}
where the factors
\begin{eqnarray}
  \label{eq:fml}
  H^l_m (\lambda) = {m\choose l} \frac{I_l(1/\lambda)
  I_{m-l}(1/\lambda)}{S_m(1/\lambda)}
\end{eqnarray}
include the combinatorial factor arising from the number of ways to choose $l$
variables $x_a$ close to one saddle point, $x_-$ say, and $m-l$ variables $x_b$ in
the vicinity of the other saddle point. Having established the asymptotic expression
(\ref{eq:z_sumsaddle}) valid for integer $m$ we consider separately the analytic
continuation $m\to\lambda$ for bosons and fermions.

\subsection{Bosonic statistics}
\label{sec:bos_stat}

To obtain the bosonic one-body density matrix one should treat $m$ as an
\textit{even integer} before taking the limit $m\to\lambda$. In this case the
main contribution to the sum (\ref{eq:z_sumsaddle}) comes from the central point
$l=m/2$. It behaves as $(N\sin(\alpha/2))^{-m^2/2\lambda}$ and substituting
$m=\lambda$ yields the result expected from the conformal field theory $G_1\sim
(N\sin(\alpha/2))^{-\lambda/2}$. To perform analytic continuation we rearrange the
sum by changing the summation index $l=m/2+k$ and letting $k$  run from $-\infty$
to $+\infty$. The coefficient of the dominant ($k=0$) term is given by $A_\lambda^2
(\lambda ) = \lim_{m\to\lambda} H^{m/2}_m$. The analytic continuation of this
expression is described in detail in Appendix~\ref{app:analAn}. The
result is
\begin{eqnarray}
  \label{eq:alambda}
  A_\lambda (\lambda) = \frac{\Gamma^{1/2} (1+\lambda)} {\Gamma
  \left(1+\lambda/2\right)}\exp\int_0^\infty \frac{dt}{t}e^{-t}
  \left(\frac{\lambda}{4}- \frac{2\left(\cosh\frac{t}{2}-1\right)}
  {\left(1-e^{-t}\right)\left(e^{t/\lambda}-1\right)}\right) .
\end{eqnarray}

The coefficients $D^2_k(\lambda) =\lim_{m\to\lambda} H^{m/2+k}_m/H^{m/2}_m$ of the
oscillating $k\neq 0$ terms are obtained straightforwardly:
\begin{eqnarray}
  \label{eq:D}
  D_k(\lambda) &=& \lim_{m\to\lambda} \prod_{a=1}^k
  \frac{\Gamma\left(\frac{m/2+a}{\lambda}\right)}{\Gamma\left(\frac{m/2+1-a}{\lambda}\right)}
  = \prod_{a=1}^k\frac{\Gamma\left(\frac{1}{2}+\frac{a}{\lambda}\right)}
  {\Gamma\left(\frac{1}{2}+\frac{1}{\lambda}
  -\frac{a}{\lambda}\right)}
\end{eqnarray}
and can be calculated recursively
\begin{eqnarray}
  D_1(\lambda)&=&\frac{\Gamma\left(\frac{1}{2}+\frac{1}{\lambda}\right)}
  {\Gamma\left(\frac{1}{2}\right)},\qquad\qquad D_{k+1}(\lambda) =
  \frac{\Gamma\left(\frac{1}{2}+\frac{1}{\lambda}+\frac{k}{\lambda}\right)}
  {\Gamma\left(\frac{1}{2}-\frac{k}{\lambda}\right)}D_k (\lambda) .
\end{eqnarray}

The correlation function is thus represented as a sum of one smooth component and
oscillatory corrections due to the short-distance correlations:
\begin{eqnarray}
  \label{eq:g1_boson}
  G^B_1 (\alpha) \sim \frac{A_\lambda^2 (\lambda)}{|2X|^{\lambda/2}}
  \left(1+2\sum_{k=1}^\infty (-1)^k D_k^2 (\lambda) \frac{\cos\Big(k
  N\alpha\Big)} {|2X|^{2k^2/\lambda}} \right).
\end{eqnarray}

Here we have changed the number of particles back to $N$, and defined $X=(N-1)\sin
\frac{\alpha}{2}$. From the last expression one sees that  if $\lambda$
 is an even integer
the infinite sum in the expression (\ref{eq:g1_boson}) becomes finite. For
$\lambda=2 n$ the correlation function contains only $n$ oscillatory components.

\subsection{Fermionic statistics}
\label{sec:ferm_stat}

To deal with fermions, $m$ has to be considered as an \textit{odd integer} prior to
the limiting procedure $m\to\lambda$. In this case the dominant contribution to the
sum (\ref{eq:z_sumsaddle}) comes from the terms $l=(m\pm 1)/2$, which being combined
together for $m\to\lambda$ lead to the oscillating behaviour $\sim
2\sin\left(\frac{1}{2}(N+1)\alpha\right)/((2N\sin(\alpha/2))^{\lambda/2+1/{2\lambda}}$
of the one-body density matrix. This behaviour is again in complete agreement with
the conformal field theory. The coefficient of this term $C^2_\lambda (\lambda) =
\lim_{m\to\lambda} H^{(m+1)/2}_m$ is calculated by an analytic continuation as
explained in  Appendix~\ref{app:analAn}. The result is
\begin{eqnarray}
  \label{eq:C}
  C(\lambda)=\Gamma^{1/2}
  (\lambda)\frac{\Gamma(1/2+1/2\lambda)}{\Gamma(1/2+\lambda/2) }
  \exp\int_0^\infty \frac{dt}{t}e^{-t}
  \left(\frac{\lambda}{4}-\frac{1}{4\lambda} + \frac{2
  e^{t/2\lambda}\left(\cosh \frac{t}{2\lambda} - \cosh \frac{t}{2}\right) }
  {\left(1-e^{-t}\right)\left(e^{t/\lambda}-1\right)}\right).
\end{eqnarray}
and the coefficients $F^2_k (\lambda) = \lim_{m\to\lambda} H^{(m+k)/2}_m
/H^{(m+1)/2}_m$ are given by
\begin{eqnarray}
  \label{eq:F}
  F_k(\lambda) &=&
  \left(\frac{\lambda^2-(2k-1)^2}{\lambda^2-1}\right)^\frac{1}{2}\;
  \prod_{a=1}^{k-1}\frac{\Gamma\left(1+\frac{1+2a}{\lambda}\right)}
  {\Gamma\left(1+\frac{1-2a}{\lambda}\right)},
\end{eqnarray}
or by the recursion relation
\begin{eqnarray}
  F_1(\lambda)&=&1,\qquad\qquad F_{k+1}(\lambda) =
  \left(\frac{\lambda^2-(2k+1)^2}{\lambda^2-(2k-1)^2}\right)^\frac{1}{2}\;
  \frac{\Gamma\left(1+\frac{1+2k}{\lambda}\right)}
  {\Gamma\left(1+\frac{1-2k}{\lambda}\right)}\; F_k (\lambda).
\end{eqnarray}

Restoring the number of particles and using $X=(N-1)\sin\frac{\alpha}{2}$, the
fermionic one-body density matrix is finally given by a sum of oscillatory terms:
\begin{eqnarray}
  \label{eq:g1_fermion}
  G^F_1 (\alpha) \sim \frac{C^2(\lambda)}{ |2X|^{\frac{\lambda}{2}+
      \frac{1}{2\lambda}-1}}\frac{1}{X} \sum_{k=1}^\infty (-1)^{k-1} F^2_k
      (\lambda) \frac{\sin\Big((k-1/2)N\alpha\Big)}{|2X|^{2k(k-1)/\lambda}} .
\end{eqnarray}
For $\lambda=2n-1$ the total number of terms in the sum is $n$.

\section{Thermodynamic limit}
\label{sec:thermo_lim}

We now take the thermodynamic limit of the above expressions by recalling that
$\alpha=2\pi x/L$ so that $X=N\sin(\pi x/L)\simeq \pi x N/L = \pi n x = k_F x$ in
the limit of infinite $N$ and $L$. Here we have also defined $k_F=\pi n$, the Fermi
momentum by analogy with free one-dimensional spinless fermions. In the thermodynamic limit
the expressions (\ref{eq:g1_boson}) and (\ref{eq:g1_fermion}) become:
\begin{eqnarray}
  \label{eq:g1_boson_thermo}
  \label{g1B}
  \frac{g^B_1 (x)}{n} &\simeq& \frac{A^2 (\lambda)}{|2 k_F x|^{\lambda/2}}
  \left(1+2\sum_{m=1}^\infty (-1)^m \frac{D_m^2 (\lambda) \cos 2m k_F x} {|2
  k_F x|^{2m^2/\lambda}} \right) \\
 \label{eq:g1_fermion_thermo}
 \label{g1F}
  \frac{g^F_1 (x)}{n} &\simeq& \frac{C^2(\lambda)}{|2k_F x|^{\frac{\lambda}{2}+
      \frac{1}{2\lambda}-1}}\frac{1}{k_F x} \sum_{m=1}^\infty (-1)^{m-1}
      \frac{F^2_m (\lambda) \sin (2m-1)k_F x }{(2 k_F x)^{2m(m-1)/\lambda}}.
\end{eqnarray}
These expressions agree completely with the results of Haldane based on the
universal hydrodynamic theory for compressible quantum fluids \cite{Haldane1981}.
The power-law universal decay of correlations with the distance provides the value
of the corresponding Luttinger parameter $K=1/\lambda$. The coefficients
$A(\lambda)$ (\ref{eq:alambda}), $D_m(\lambda)$ (\ref{eq:D}) for bosons and
$C(\lambda)$ (\ref{eq:C}), $F_m (\lambda)$ (\ref{eq:F}) for fermions, are
model-specific non-universal numbers. They are strikingly similar to the
analogous coefficients which appear in the correlation functions of the
Heisenberg spin chain \cite{Lukyanov_etal}. Here we derive them for the first time
in the case of the Calogero-Sutherland model.

{}From the expressions (\ref{eq:g1_boson_thermo}), (\ref{eq:g1_fermion_thermo}) we
extract the singular behaviour of the momentum distribution using the definition
(\ref{nk_def}). For bosons, the leading term in (\ref{eq:g1_boson_thermo}) yields
the leading divergence for small momenta:
\begin{eqnarray}
  \label{eq:nk_boson_thermo}
  \label{nkb:infrared}
  n_k \simeq \frac{2^\alpha }{\pi}A^2(\lambda)
  \Gamma\left(\alpha\right)\cos\frac{\pi\alpha}{2}
  \left|\frac{k}{k_F}\right|^{-\alpha},
\end{eqnarray}
where the critical exponent $\alpha=\alpha(\lambda)$ is defined as
$\alpha=1-\lambda/2$. This result is valid for $0<\lambda<2$, so that $0<\alpha<1$.
For the special value of $\lambda=1$, which corresponds to the system of
impenetrable bosons \cite{Girardeau1960} we have $\alpha=1/2$. For the value
$\lambda=1/2$ Sutherland \cite{Sutherland1992} was using a numerical estimate
to suggest $\alpha=1/\sqrt{2}=0.707\ldots$. We report here the exact value $\alpha=3/4$. For
$\lambda=2$ the critical exponent $\alpha=0$ is consistent with the logarithmic
divergence found in \cite{Sutherland1971}. The oscillating terms in
(\ref{eq:g1_boson_thermo}) contribute to the weaker singularities at the points
$k=\pm 2k_F,\pm 4k_F,\ldots$.

For fermions, due to the oscillating character of the one-body density matrix the
dominant singularities appear at $k=\pm k_F$. The leading behaviour is extracted
from the first term in (\ref{eq:g1_fermion_thermo}). For $k\to k_F$ we have
\begin{eqnarray}
  \label{eq:nk_fermion_thermo}
  n_k-\frac{1}{2}\simeq \frac{2^{-\beta} C^2(\lambda)}{\pi\beta}
    \Gamma\left(1-\beta \right)\cos \left(\frac{\pi(\beta-1)}{2}\right)
    \left|\frac{k-k_F}{k_F}\right|^\beta \sgn (k_F-k)
\end{eqnarray}
and similarly for $k\to -k_F$. The critical exponent is defined as $\beta
=\beta(\lambda) =\beta(1/\lambda) = \lambda/2+1/2\lambda -1$, so that  the power
law behaviour of the momentum distribution is the same for $\lambda$ and
$1/\lambda$. The result (\ref{eq:nk_fermion_thermo}) is valid for $0<\beta<1$ which
implies $2-\sqrt{3}<\lambda < 2+\sqrt{3}$. For $\lambda=1$ we have free fermions and
$\beta(1) = 0$ corresponding to the Fermi-Dirac distribution. In other the special
cases we have $\beta(1/2)=\beta(2) = 1/4$. Additional singularities exist at $k=\pm
3k_F,\pm 5k_F ,\ldots$.

\section{Short-distance correlations of the Calogero-Sutherland model}
\label{sec:shortcorr}

The results in the previous section provide the long-distance behaviour of the
one-body density matrix for the Calogero-Sutherland model and consequently the
small-momenta behaviour of its momentum distribution. It is also possible to extract
the short-distance properties of this model. We use the method introduced by
Olshanii and Dunjko \cite{Olshanii2003} to relate the tails of the momentum
distribution to singularities of the wave function. To present this method we
consider bosonic statistics and discuss later the corresponding modifications for
fermions.

The momentum distribution (\ref{nk_def}) is rewritten in the following form:
\begin{eqnarray}
  \label{eq:nk_wave}
  n_k=n\int dx_2\ldots dx_N |\Phi(k,x_2,\ldots,x_N)|^2,
\end{eqnarray}
where we have defined the Fourier transform of the ground state wave function
(\ref{gs}) with respect to its first coordinate:
\begin{eqnarray}
  \label{eq:wave_fourrier}
  \Phi(k,x_2,\ldots,x_N) &=& \int_0^L dx e^{-ikx} \Phi(x,
  x_2,\ldots,x_N)\nonumber \\ &=&C_N(\lambda) \prod_{2\le k<l\le N}
  \left|e^{2\pi x_k/L}-e^{2\pi x_l/L}\right|^\lambda \int_0^L dx e^{-ikx}
  \prod_{l=2}^N \left|e^{2\pi x /L}-e^{2\pi x_l/L}\right|^\lambda .
\end{eqnarray}
The remaining integral can be evaluated using the following property of Fourier
transforms. Let $f(z)$ have a singularity of the form $f(z) = |z-z_0|^\alpha g(z)$ where
$g(z)$ is regular at $z=z_0$ and $\alpha>-1$, $a\neq 0,2,4,\ldots$. Then
\begin{eqnarray}
  \label{eq:fourrier_property}
  \int_{-\infty}^{+\infty} dz e^{-ipz} f(z) =
  2\cos\left(\frac{\pi(\alpha+1)}{2}\right) \Gamma(1+\alpha)
  \frac{g(z_0)e^{-ipz_0}}{|p|^{\alpha+1}}+O\left(\frac{1}{p^{\alpha+2}}\right).
\end{eqnarray}
If $f(z)$ has several singularities, then the right hand side equals the sum of the
corresponding contributions. Using this fact and (\ref{eq:fourrier_property}) the
large-$k$ behaviour of the wave function (\ref{eq:wave_fourrier}) is determined by
the singularities at the positions of the remaining particles:
\begin{eqnarray}
  \label{eq:wave_fourrier_res}
  \Phi(k,x_2,\ldots, x_N) &\simeq& \frac{2}{|k|^{\lambda+1}}\left(\frac{2\pi}{
  L}\right)^\lambda \cos\left(\frac{\pi(\lambda+1)}{2}\right)
  \Gamma(1+\lambda) C_N(\lambda) \prod_{2\le k<l\le N} \left|e^{2\pi
  x_k/L}-e^{2\pi x_l/L}\right|^\lambda\nonumber\\ &\times&\sum_{j=2}^N
  e^{-ikx_j} \prod_{l\neq j} \left|e^{2\pi x_j/L}-e^{2\pi
  x_l/L}\right|^\lambda .
\end{eqnarray}
Substituting this expansion into (\ref{eq:nk_wave}) and keeping only $k$-independent
diagonal terms in the double sum leads to the following result for the asymptotic
behaviour of the momentum distribution:
\begin{eqnarray}
  \label{eq:nk_wav_res}
  n_k \simeq K_N (\lambda) \left|\frac{k_F}{k}\right|^{2+2\lambda},
\end{eqnarray}
with the constant $K_N (\lambda)$ defined by the following average in the ground
state of $N-1$ particles:
\begin{eqnarray}
  \label{eq:KN}
  K_N (\lambda)&=&
  \frac{2^{2+2\lambda}}{\pi^2}\cos^2\left(\frac{\pi(\lambda+1)}{2}\right)
  \frac{\Gamma^3(1+\lambda) \Gamma(1-\lambda+\lambda N)}
  {N^{2\lambda}\Gamma(1+\lambda N)} \nonumber\\ &\times& \int dx_2\ldots dx_N
  \left|\Phi(x_2,\ldots,x_N)\right|^2 \prod_{l=3}^N \left|e^{2\pi
  x_2/L}-e^{2\pi x_l/L}\right|^{2\lambda} .
\end{eqnarray}
The result (\ref{eq:nk_wav_res}) is valid also for fermions for $\lambda\neq
1,3,\ldots$, with a modification of the proportionality constant:
\begin{eqnarray}
  \label{eq:KNferm}
  \widetilde{K}_N(\lambda) =\tan^2\left(\frac{\pi(\lambda+1)}{2}\right)
  K_N(\lambda).
\end{eqnarray}

The result (\ref{eq:nk_wav_res}) suggests the following short range expansion for the
one-body density matrix as a sum of analytic and non-analytic functions:
\begin{eqnarray}
  \label{eq:g1_short}
  \label{SR}
  \frac{g_1(x)}{n} &=& 1+c_1 (\lambda) k_F x +\frac{c_2 (\lambda)}{2!} (k_F x)^2 +\frac{c_3
    (\lambda)}{3!}(k_F x)^3 +\ldots \nonumber \\
  &+& a(\lambda) |k_Fx|^{1+2\lambda}+O\left(|k_F x|^{2+2\lambda}\right) .
  %g_1(x)/n &=& 1+c_1 (\lambda)k_F x +\frac{c_2 (\lambda)}{2!} (k_F x)^2
  %+\frac{c_3 (\lambda)}{3!}(k_F x)^3
  %%%+\frac{c_4(\lambda)}{4!}  (k_F x)^4
  %+\ldots \nonumber \\
  %&+& a(\lambda) |k_F x|^{1+2\lambda}+O\left(|k_F x|^{2+2\lambda}\right)
\end{eqnarray}
The coefficients $c_{l}(\lambda)$ in the Taylor expansion of the analytic part of
$g_1$ are the corresponding moments of the momentum distribution
\begin{eqnarray}
  \label{eq:nk_moments}
  %c_{l}(\lambda) = \frac{i^l}{k_F^{l}} \int \frac{dk}{2\pi n} k^{l} n_k .
  c_{l}(\lambda) = \frac{i^l}{n^l} \int \frac{dk}{2\pi n} k^l n_k .
\end{eqnarray}
Due to the time-reversal symmetry the momentum distribution is an even function,
$g_1(x)$ is real and odd moments vanish. The non-analytic part of the one-body density
matrix starts as $|x|^{2\lambda+1}$ with the coefficient which can be related to the
high-momentum tails (\ref{eq:nk_wav_res}) using (\ref{eq:fourrier_property}):
\begin{eqnarray}
  \label{eq:a_lambda}
  a(\lambda) =
  %\frac{\pi}{2} \frac{K_N(\lambda)}{\cos(\pi(1+\lambda))\Gamma(2+2\lambda)}
  \frac{\pi K_N(\lambda)}{2\cos(\pi(1+\lambda))\Gamma(2+2\lambda)} .
\end{eqnarray}
Non-analyticity of the density matrix at $x=0$ reflects the fact seen already from
(\ref{eq:nk_wav_res}) that the $l$-th moment $c_l(\lambda)$ diverges for
$l>2\lambda+1$.

Provided they exist, the moments $c_{2l}$ are the same for bosons and fermions,
a fact noticed by Sutherland in \cite{Sutherland1992}. Therefore for sufficiently
large $\lambda$ the fermionic and bosonic one-body density matrices have the same
leading Taylor expansion. This fact is easily explained on physical grounds,
noticing that the strong repulsion prevents all exchange effects and particles do
not ``feel'' the quantum statistics.

The second moment $c_2(\lambda)$ can be calculated explicitly since it is given by
minus the kinetic energy per particle (in units of the Fermi energy). To obtain it one
uses the Hellman-Feynman theorem to extract the potential energy from the dependence of
the total energy on the strength of the particle-particle interaction. The ground state
energy of the Calogero-Sutherland Hamiltonian (\ref{ham}) is known exactly
\cite{Sutherland1971} and is equal  to
\begin{eqnarray}
\frac{E}{N}=\frac{\left\langle H\right\rangle}{N}
=\frac{\pi^2\lambda^2}{6}\frac{\hbar ^{2}n^{2}}{m} .
\label{E}
\end{eqnarray}

We note that the potential energy is linear in $g=\lambda(\lambda-1)$ and can be
obtained by differentiating  the ground state energy:
\begin{eqnarray}
\frac{E_{pot}}{N}=\frac{g}{N} \left\langle \frac{\partial}{\partial
    g}H\right\rangle =\frac{g}{N}\frac{\partial}{\partial g}\left\langle
H\right\rangle =\frac{\pi^2\lambda^2(\lambda-1)}{3(2\lambda-1)}\frac{\hbar
^{2}n^{2}}{m}, \qquad\lambda>1/2 .
\label{Epot}
\end{eqnarray}
The kinetic energy is then obtained as a difference of the total ground state energy
(\ref{E}) and potential energy (\ref{Epot}):
\begin{equation}
\frac{E_{kin}}{N}=\frac{E-E_{pot}}{N}=\frac{\pi^2\lambda^2}{6(2\lambda-1)}
\frac{\hbar^2n^2}{m},\qquad \lambda>1/2  .
\label{Ekin}
\end{equation}
Finally, for the second moment $c_2$ we obtain the following expression:
\begin{equation}
c_2=-\frac{\lambda^2}{3(2\lambda-1)}  .
\label{c2}
\end{equation}

\begin{figure}
\begin{center}
\includegraphics[width=0.5\columnwidth,angle=-90]{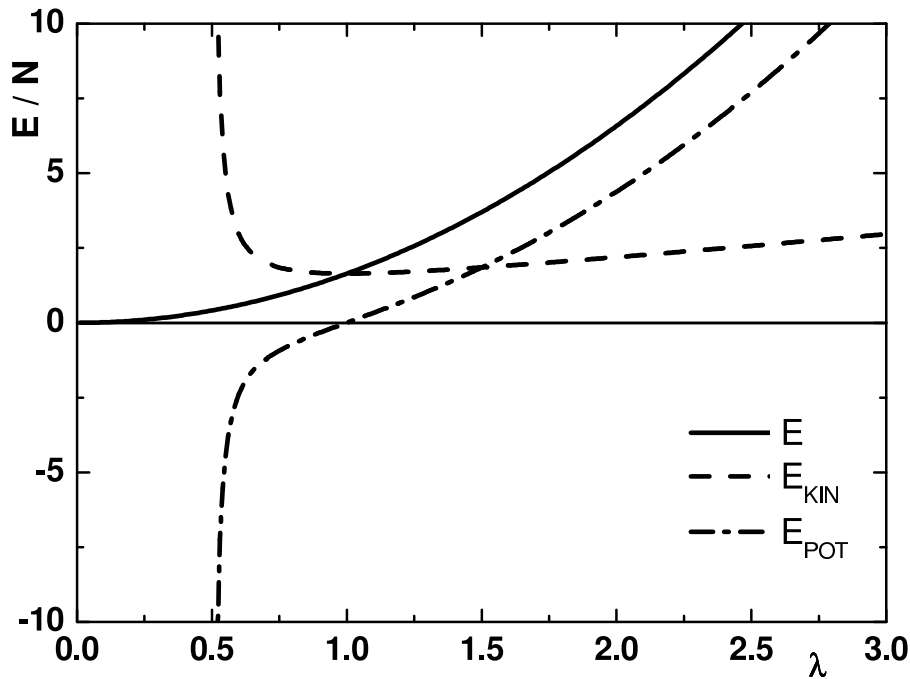}
\caption{Ground state energy of the CSM as a function of the interaction
parameter $\lambda$: solid line -- total energy per particle (\ref{E}), dashed line
-- kinetic energy per particle (\ref{Ekin}), dot-dashed line -- potential energy per
particle (\ref{Epot}). Energies are measured in units of $\hbar^2n^2/m$.}
\label{Fig:E}
\end{center}
\end{figure}

We plot the total, kinetic and potential energies in Fig.~\ref{Fig:E}. The total energy
is always positive and the compressibility also remains positive so that the system
is (thermo-) dynamically stable. The case $\lambda>1$ corresponds to repulsion
between fermions while for $\lambda<1$ the interaction between fermions is
attractive. The ground state energy is independent of statistics, yet the
interpretation for bosons is more involved: the ground state wave function
(\ref{gs_mod}) always describes repulsion between particles, despite the negative
value of the potential energy for $\lambda<1$. This paradox is due to the singular
character of interactions, namely the requirement that wave functions are zero for
coinciding positions of the particles.

As can be seen in Fig.~\ref{Fig:E} both the kinetic and potential energies diverge as
$\lambda\to 1/2$. It corresponds to the critical value of the coupling constant
$g=\lambda(\lambda-1)=-1/4$, below which particles fall to the center
\cite{LandauLifshitzIII,Sutherland1971}. For $\lambda<1/2$ the expression
(\ref{Ekin}) is not valid, since it predicts negative expectation value for the kinetic
energy. The same applies to the potential energy (\ref{Epot}). However the
total energy is finite and analytic in the full range $\lambda>0$ as follows
from Eq.~(\ref{E}). Approaching the value $\lambda=1/2$ from above the
divergences of kinetic and potential energies cancel each other.
In the interval $0<\lambda<1/2$  one cannot evaluate separately the potential
energy using the Hellmann-Feynman theorem (\ref{Epot}) due to the singular character of
the ground-state wave function in thi regime. To the best of our knowledge, this
intriguing behaviour has not been noticed in the literature on the
Calogero-Sutherland  model despite the simple analysis based on use of
the Hellmann-Feynman theorem.

\section{Numerical results and discussion }
\label{sec:num_res_disc}

\subsection{Monte Carlo method}
\label{sec:monte_carlo}

Quantum Monte Carlo methods have been successfully applied  to the
investigation of the
equation of state and correlation functions in a number of one-dimensional systems
\cite{MC_LiebLiniger, MC_SuperTonks,MC_1DDipoles}. We resort to the Quantum Monte Carlo
technique in order to evaluate multidimensional integrals
Eqs.~(\ref{g1_def},\ref{g2_def}) numerically. This calculation of the correlation
functions is exact in a statistical sense, the small statistical uncertainty can be
reduced by increasing the length of the simulation runs. An advantage of the CSM is
that the ground state wave function is known exactly and can be written in a simple
explicit way, as given by Eq.~(\ref{gs_mod}), thus facilitating the calculations at
zero temperature. Another important advantage of the CSM is related to the ``sign
problem'' of fermionic Monte Carlo simulations. Here the permutation term can be
factorized leading to a much simpler and efficient code than in three-dimensional
systems, where the symmetrization of the fermionic wave function leads to the evaluation
of Slater determinants. We use the Metropolis \cite{Metropolis53} algorithm for sampling
the square of the wave function and generating  a series of states (Markovian chain)
having the desired probability distribution. The correlation functions are then
calculated as averages over the Markovian chains. The calculation of the one-body
density matrix $g_1(x)$ is performed by displacing a certain particle (let us choose
$x_1$ as an example) by a distance $x$, so that $x_1'=x_1+x$ and averaging the ratio
$\Phi(x_1',x_2,...)/\Phi(x_1,x_2,...)=s_{B,F} \prod_{i\ne 1}
|\sin(x_1'-x_i)/\sin(x_1-x_i)|^\lambda $ where $s_B=1$ and $s_F=\prod_{i\ne
1}\sgn(x_1'-x_i)/\sgn(x_1-x_i)$. The average is performed with the probability
distribution $|\Phi(x_1,...,x_N)|^2$. Similarly, we accumulate the pair distribution
function by following the Markovian chain generated by the Metropolis algorithm and
measuring the interparticle distance. The momentum distribution and static structure
factor are calculated by  means of  Fourier transforms as defined by
Eqs.~(\ref{nk_def},\ref{sk_def}). In what follows we present our results for the
correlation functions in different interaction regimes.

\subsection{One body density matrix}
\label{sec:num_g1}

\begin{figure}[ht]
\begin{center}
\includegraphics[width=0.78\columnwidth]{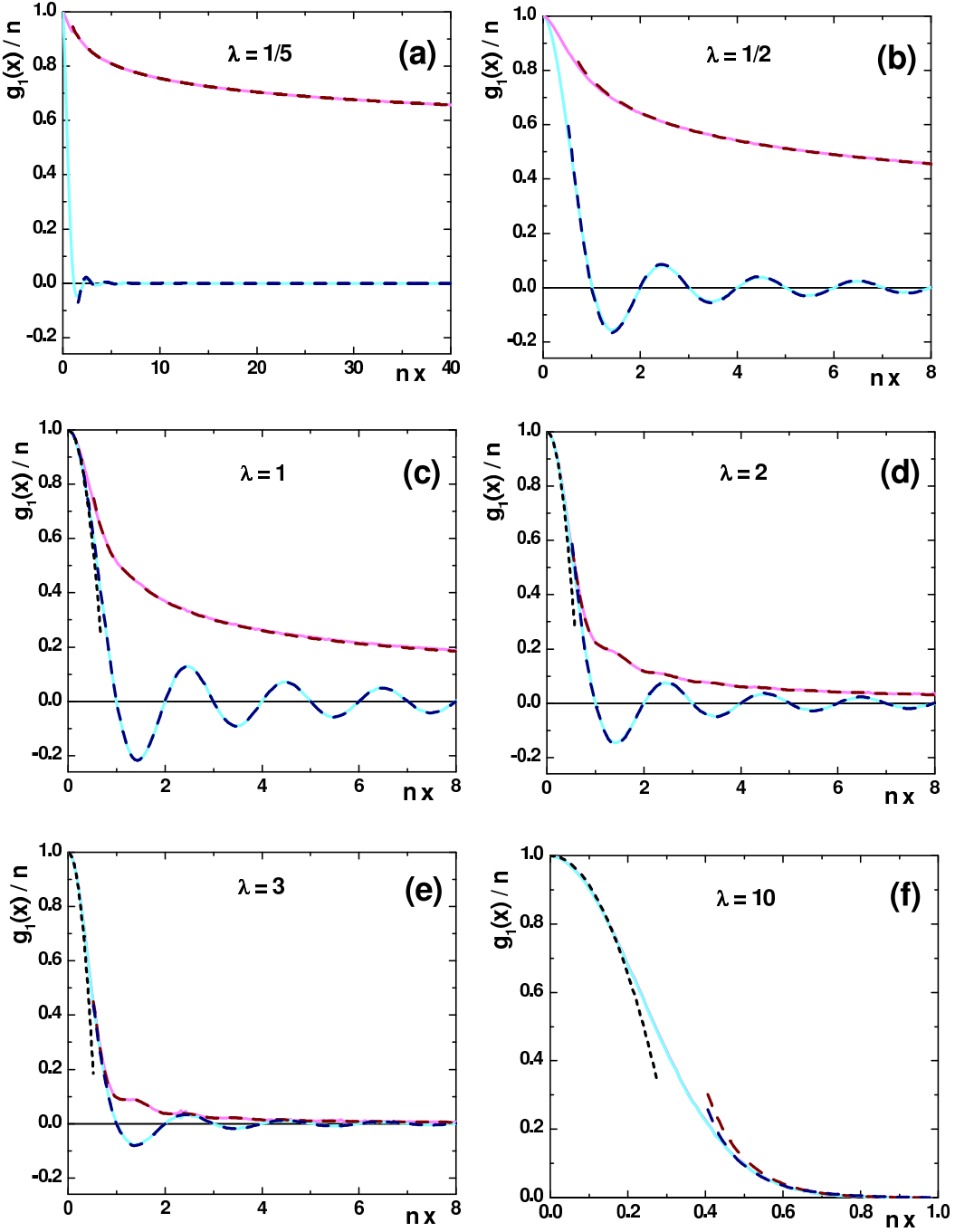}
\caption{(Color online) One-body density matrix $g_1(x)$ in the thermodynamic limit for different
values of the interaction parameter $\lambda$. Solid lines: bosons (upper line);
fermions (lower line). Long dashed lines: long range expansion for bosons,
Eq.~(\ref{eq:g1_boson_thermo}), (upper line); for fermions,
Eq.~(\ref{eq:g1_fermion_thermo}), (lower line). Short dashed line (c,d,e,f):
short-range expansion Eqs.~(\ref{eq:g1_short},\ref{c2}). Note that the fermionic
long-range expansion, Eq.~(\ref{eq:g1_fermion_thermo}), is exact for $\lambda=1$
(c).}
\label{Fig:g1}
\end{center}
\end{figure}

The results for the one-body density matrix are presented in Figs.~\ref{Fig:g1} for
Bose-Einstein and Fermi-Dirac statistics. Comparing the numerical results to
Eqs.~(\ref{eq:g1_boson_thermo}) and (\ref{eq:g1_fermion_thermo}) we find that the
asymptotic expansion works extremely well even for distances of the order of mean
interparticle separation. It is important to note that the series
(\ref{eq:g1_fermion_thermo}) and (\ref{eq:g1_boson_thermo}) are asymptotic rather
than convergent. The larger  the distance $x$ the more terms should be
summed, while small distances are well described with only  a few terms in the sum.

It is easy to see from Eq.~(\ref{eq:g1_boson_thermo}) that the off-diagonal
long-range order is absent as $g_1(x)$ always vanishes for large $x$. Still, for
bosons and for small values of $\lambda$ the leading term $g_1(x)\sim A^2/|2\pi
x|^{\lambda/2}$ has a slow power-law decay, which is a manifestation of a quasi
off-diagonal long-range order or {\it quasi-condensation}. This behaviour is shown
in Figs.~\ref{Fig:g1}a-\ref{Fig:g1}b for small values of $\lambda$ corresponding to
weak interaction between particles. The smaller  $\lambda$  is, the more
pronounced is
the presence of a quasi-condensate. In this regime the one-body density matrix
remains significantly different from zero even at distances much larger than the
mean interparticle distance. The slowly decaying off-diagonal correlation in this
regime  is well described by the dominant term in the
expansion~(\ref{eq:g1_boson_thermo}). Oscillating corrections, corresponding to
$m\ge 1$ terms in Eq.~(\ref{eq:g1_boson_thermo}), decay rapidly on the scale of
a few interparticle separations and are unimportant in this regime.

In this regime the off-diagonal correlations in fermionic systems are qualitatively
different. For fermions, the one-body density matrix drops quickly from unity and
oscillates around zero. In the fermionic case the main contribution to the long
range asymptotics comes from the leading oscillating $m=1$ term,
Eq.~(\ref{eq:g1_fermion_thermo}).

With increasing $\lambda$ the oscillating $m>0$ terms in the long-range expansion of
the bosonic one-body density matrix, Eq.~(\ref{eq:g1_boson_thermo}), become
important. Those contributions introduce oscillations corresponding to momenta
$2mk_F, m=1,2,...$, where $k_F=\pi n$ is the Fermi momentum. This behaviour can be
attributed to the short-range order in the Calogero-Sutherland model, which becomes
more important as the interaction strength increases. The results for intermediate
values of interactions are depicted in Fig.~\ref{Fig:g1}b-\ref{Fig:g1}e.

For a bosonic system  these short-range correlations are a manifestation of
``fermionization''. Indeed, for $\lambda=1$ the bosonic CSM becomes equivalent to
 a system of zero-range impenetrable bosons having the density
correlations of free
fermions. The correlation functions involving the phase are, however, drastically
different and  were the subject of the classic works  of
Refs.~\cite{Lenard1964,VaidyaTracy1979}. Recently there  has been a revival of
interest in this
model \cite{TG:theory,Olshanii2003,MC_LiebLiniger,Gangardt2004} due to realization
of impenetrable bosons in several experiments with cold atoms \cite{TG:experiments}.
For the fermionic CSM at $\lambda=1$, the one-body density matrix is given by a
standard expression for one-dimensional spinless fermions:
\begin{eqnarray}
\frac{g_1^F(x)}{n}=\frac{\sin k_Fx}{k_Fx}, \qquad\lambda=1.
\label{g1Flambda1}
\end{eqnarray}
It is interesting to note that the asymptotic expansion
(\ref{eq:g1_fermion_thermo}) is exact for $\lambda=1$. The results for the one-body
density matrix at $\lambda=1$ are shown in Fig.~\ref{Fig:g1}b both for free fermions
and impenetrable bosons.

In the regime $\lambda>1$ the system enters the quasi-crystal regime. The off-diagonal
decay of the one-body density matrix is greatly enhanced. The oscillating terms in
Eq.~(\ref{g1B}) become relevant and oscillations in $g_1^B(x)$ start to be visible
(see Figs.~\ref{Fig:g1}d-\ref{Fig:g1}e) which reflects the appearance of the short
range quasi-crystal order. For $\lambda = 2$ the bosonic one-body density matrix is
known exactly \cite{Sutherland1971} and is given by the expression
\begin{eqnarray}
\frac{g_1^B(x)}{n} = \frac{\Si (2k_F x)}{2 k_F x}, \qquad\lambda=2,
\label{g1Blambda2}
\end{eqnarray}
where $\Si(x)$ is the sine integral function. The expression (\ref{g1B}) gives
$g_1^B/n=\pi/4k_F x-\cos 2 k_F x\,/\,4 k_F^2 x^2$ which coincides with the large $x$
expansion of (\ref{g1Blambda2}). We note that this behaviour of $g_1^B(x)$ is
critical between long-range and short-range correlations, in the sense that the integral
of $g_1^B(x)$ over  space diverges for $\lambda<2$ and converges for $\lambda>2$.
The description in terms of a quasi-condensate is applicable only in the
long-range regime $\lambda<2$.

For extremely strong interactions (for example, $\lambda=10$, see Fig.~\ref{Fig:E})
the potential energy dominates the total energy and quantum fluctuations are
suppressed. In this regime the role of quantum statistics becomes irrelevant. This
is due to the onset of the quasi-crystalline order, in which particles
form a local crystal lattice so that the exchange effects of quantum
statistics are irrelevant. As a result the one-body density matrices for bosons and fermions are
very similar (see Fig.~\ref{Fig:g1}f). One finds that the fermionic one-body density
matrix is positive in a large region and vanishes otherwise. The short-range
expansion (\ref{SR}) describe quite well both fermionic and bosonic one-body density
matrices. The quasi-crystal order is better probed by density correlations described
in the subsection~\ref{num:pair_dist}.

\begin{figure}
\begin{center}
\includegraphics[width=\columnwidth]{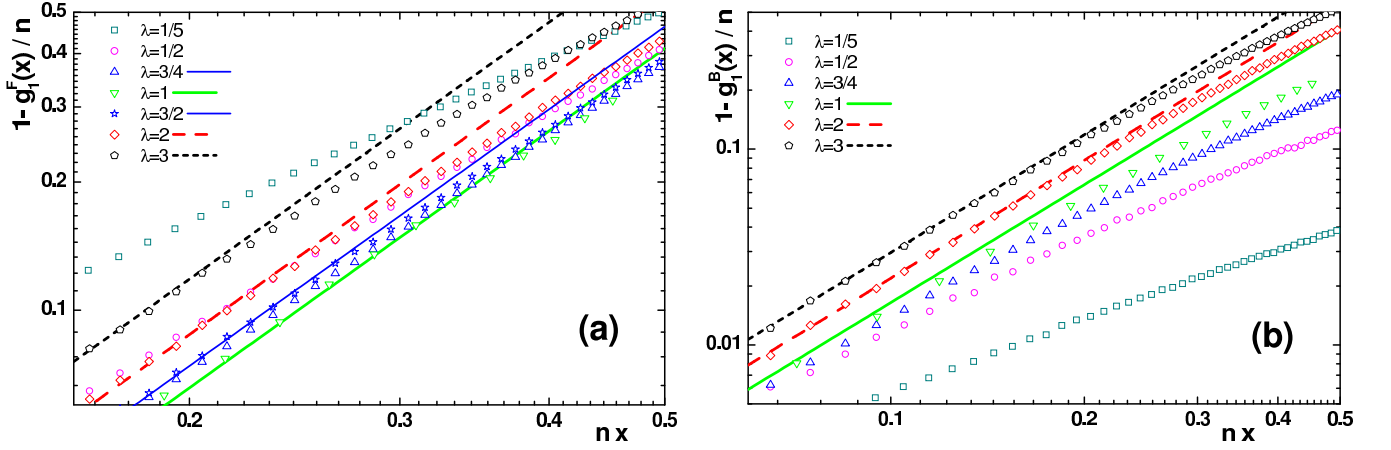}
\caption{(Color online) Short-range behavior of the bosonic (a) and fermionic (b) one-body
density matrix $g_1(x)$ in the thermodynamic limit for different values of the
interaction parameter $\lambda$. Symbols: results of the Monte Carlo simulations,
lines: analytic part of the short-range expansion ($\lambda>1$),
Eqs.~(\ref{eq:g1_short},\ref{c2}).}
\label{Fig:g1SR}
\end{center}
\end{figure}

We have investigated the short-range behaviour of $g_1(x)$ numerically and we found
a good agreement with the expression (\ref{SR}). The results are presented in
Fig.~\ref{Fig:g1SR}. This figure shows the validity of the short-range expansion for
different values of $\lambda$ and allows to find the region of applicability of Eqs.
(\ref{eq:g1_short},\ref{c2}). We see that the analytic short-range expansion holds
in a larger range in the case of fermionic statistics as the omitted higher order
terms of the short-range expansion are smaller in this case.

We point out that the coefficient $c_2$ in the short-range Taylor expansion
(\ref{SR}) has a minimum for $\lambda=1$. This is clearly seen from numerical evaluation of the
fermionic one-body density matrix presented in Fig.~\ref{Fig:g1SR}a. Interestingly,
there are pairs of $\lambda$ (for example, $\lambda$=3/4 and $\lambda$=3/2) with the
same $c_2$. The numerical results of the short-range behavior of a bosonic one-body
density matrix are presented in Fig.~\ref{Fig:g1SR}b. The dominant $c_2$ term in the
short-range expansion in this case is clearly seen for $\lambda=1,2,3$, while for
smaller values of $\lambda$ the nonanalytic correction becomes comparable to the
analytic contribution in the considered range $0.06<nx<0.5$.

We estimate the leading non analytic term and find that it goes as
$|x|^{1+2\lambda}$ both for bosons and fermions. In the regime $\lambda<1/2$ the
non-analytic part in Eq.~(\ref{eq:g1_short}) provides the leading contribution
$1-g_1(z)/n \propto a(\lambda)|z|^{1+2\lambda}$. The point $\lambda=1/2$ is very
special. Indeed, the kinetic energy (\ref{Ekin}) diverges at this point (see
Fig.~\ref{E}), thus the Taylor coefficient $c_2$ is also divergent. However, this
divergence is compensated by the divergence in the non-analytical term $a(\lambda)
|x|^{1+2\lambda}$, which for $\lambda=1/2$ is of the same order. We note that for
$\lambda=1$ the power of the non-analytical term becomes integer again and leads to
a cubic correction, as was obtained by Olshanii {\it et al.}~\cite{Olshanii2003}.
We prove numerically the presence of the non-analytic term proportional to
$|x|^{1+2\lambda}$. The coefficient of this term can  in principal  be
obtained from  a best fit to the numerical data. We note that while the
decay law  is the same in bosonic and fermionic systems, the coefficient
$a(\lambda)$  depends on the statistics and this leads to drastically
different behaviour (e.g. $\lambda=1/5$ in Figs.~\ref{Fig:g1SR}).
The short-range non-analytical behaviour is related to the
high momenta tails of the momentum distribution discussed in the next subsection.

\subsection{Momentum distribution}

\begin{figure}
\begin{center}
\includegraphics[width=\columnwidth]{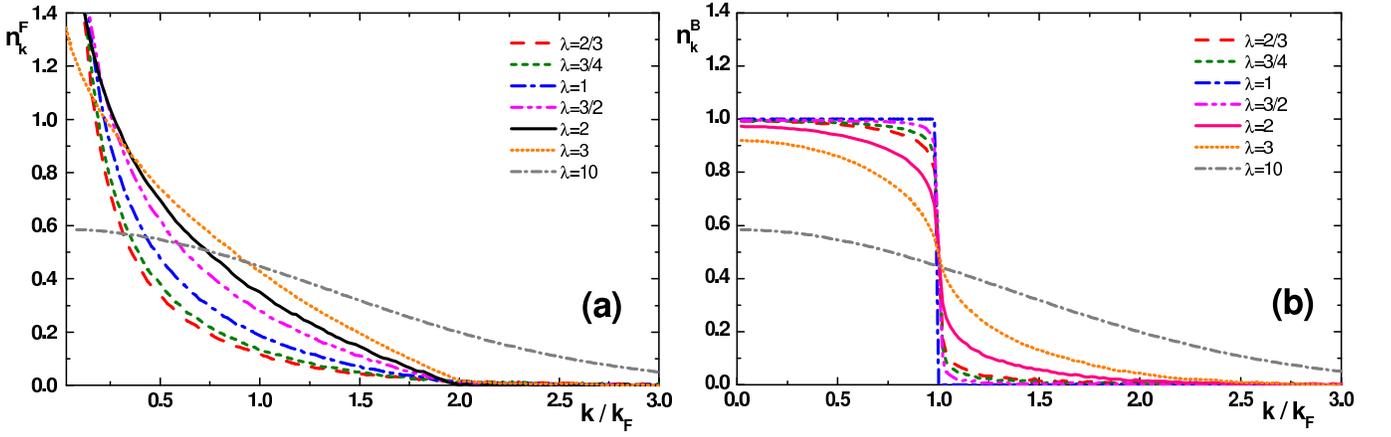}
\caption{(Color online) Momentum distributions for bosonic (a) and fermionic
(b) CSM for different values of the interaction strength. Lines: in descending value
at $k=1.5k_F$: $\lambda=10;3;2;3/2;1;3/4;2/3$.}
\label{Fig:nk}
\end{center}
\end{figure}

The Fourier transformation (\ref{nk}) relates the one-body density matrix to the
momentum distribution. The numerical results for this quantity are presented in
Fig.~\ref{Fig:nk} for the cases of bosonic and fermionic statistics.

In a system of weakly-interacting bosons, the power law decay of the one-body
density matrix results in the divergence of the momentum distribution for small
values of momenta: $n_k$ is proportional to $|k|^{1-\lambda/2}$ as it follows from
Eq.~(\ref{nkb:infrared}). This infrared divergence for small $\lambda$ is
reminiscent of  Bose-Einstein condensation, \textit{i.e.}
macroscopic occupation
of the zero momentum state. The infrared divergence is present in the {\it
quasi-condensate regime} for $\lambda<2$. At $\lambda=2$ the infrared divergence
becomes  logarithmic. From (\ref{g1Blambda2}) we have an exact result
\cite{Sutherland1971} for the momentum distribution:
\begin{eqnarray}
n^B_k =\left\{
  \begin{array}{ll}
    \frac{1}{2}\ln\frac{2k_f}{|k|}, & |k|<2k_f \\ 0, & |k|\ge 2k_f \\
  \end{array}
  \right.  ,\qquad \lambda=2.
\label{nkblambda2}
\end{eqnarray}
We note that the logarithmic divergence for $\lambda=2$ separates the power-law
divergence for $\lambda<2$ and a regular behaviour for $\lambda>2$, so that
$\lambda=2$ is the critical value beyond which the quasi-condensation disappears.

The momentum distribution for a fermionic CSM is presented in Fig.~\ref{Fig:nk}(a). In
the non-interacting case $\lambda=1$ the momentum distribution of fermions is given
by the step function at $k=\pm k_F$. For other values of $\lambda$ the steps are absent and the
momentum distribution has a power-law behaviour close to $\pm k_F$ described by
Eq.~(\ref{eq:nk_fermion_thermo}). This is a characteristic of a Luttinger liquid
behaviour present in the fermionic CSM. For very large $\lambda$ the interactions
completely destroy  the Fermi surface and the momentum
distribution is a decaying featureless function.

The behaviour of the momentum distribution for large momenta contains information
about the physics  at small length-scales. For instance, as
we see from  Eq.~(\ref{nkblambda2}), the
momentum distribution of bosons at $\lambda=2$ has a cusp at $k=\pm 2k_F$. This
discontinuity is a consequence of the oscillating term in the long-distance
asymptotics, Eq.~(\ref{eq:g1_boson_thermo}) or (\ref{g1Blambda2}) and is a
manifestation of the {\it short-range order}. Even for $\lambda<2$ the presence of
the short-range order shows itself in weaker singularities of the momentum
distribution for even multiples of $k_F$ for bosons and odd multiples of $k_F$ for
fermions.

Apart from the special integer values of the interaction parameter $\lambda$ (even
for bosons, odd for fermions), the momentum distribution has a non-analytic high
momenta tails decaying as $1/k^{2(1+\lambda)}$. At the special value $\lambda=1/2$
the momentum distribution becomes broad, with tails decaying as
$1/|k|^3$, which leads
to  a divergent mean kinetic energy $\propto\int k^2 n_k\; dk$ for $\lambda<1/2$. In
the case of zero-range impenetrable bosons ($\lambda=1$) the
Eq.~(\ref{nkb:infrared}) yields a $1/k^4$ ultraviolet behavior and has been discussed
in \cite{Olshanii2003}.

The short-range order already present the one-body density matrix and the momentum
distribution becomes evident in density correlations, which we consider in the next
section.

\subsection{Pair distribution function}
\label{num:pair_dist}

The short-range order is best probed by calculating the pair distribution function
$g_2(x)$ defined in Eq.~(\ref{g2_def}). It gives the probability of finding two
particle separated by a distance $x$. This function involves the absolute value of
the ground state function and therefore is identical for bosons and fermions. The
inverse square potential of the Calogero-Sutherland model prevents two
particles from occupying the same position, so that $g_2(0)=0$. For large interparticle
separation the
density correlations decouple and we have a general result $g_2(x)\to n^2,
|x|\to\infty$. The pair distribution function has been studied in
\cite{GangardtKamenev2001}. For distances larger than the mean interparticle
separation the following result has been obtained
\begin{eqnarray}
  \label{eq:g2_replica}
  \frac{g_2(x)}{n^2} = 1 - \frac{1}{2 \lambda (k_Fx)^2} +
2\sum\limits_{m=1}^\infty
\frac{d^2_m(\lambda)}{  (2 k_F x)^{2m^2/\lambda} } \,
\cos(2 m k_F x) ,
\end{eqnarray}
where
\begin{eqnarray}
  \label{eq:coeff_dl}
  d_l(\lambda) =
\frac{ \prod\limits_{a=1}^l \Gamma(1+a/\lambda)}
{\prod\limits_{a=1}^{l-1} \Gamma(1-a/\lambda)} =
\Gamma\((1+ {l\over\lambda}\))
\prod\limits_{a=1}^{l-1}\(({a\over \pi \lambda}\))
\sin\(({\pi a \over \lambda}\))\, \Gamma^2\(({a\over\lambda}\))    .
\end{eqnarray}

\begin{figure}
\begin{center}
\includegraphics[width=0.5\columnwidth,angle=-90]{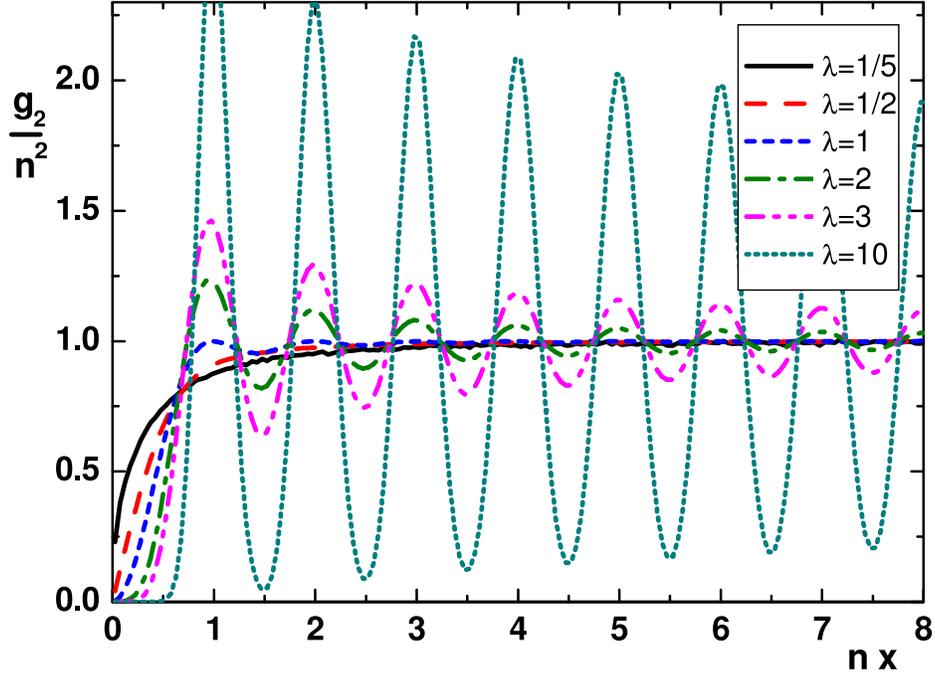}
\caption{(Color online) Pair distribution function for different values of the interaction
strength $\lambda$. Lines: in descending order of the value at the first peak:
$\lambda=10;3;2;1;1/2,1/5$.}
\label{Fig:g2}
\end{center}
\end{figure}

We have calculated the pair distribution function by using the Monte Carlo method
and results are presented in Fig.~\ref{Fig:g2}. One sees that for small values of
$\lambda$ the pair distribution function goes smoothly from zero at short distances
to the bulk constant at large distances. This is characteristic for a weakly
interacting one-dimensional Bose gas or liquid \cite{MC_LiebLiniger}. As the regime
of free fermions ($\lambda=1$) is approached some oscillations become visible in the
pair distribution. For this value of interactions the expression
(\ref{eq:g2_replica}) gives the exact result:
\begin{eqnarray}
\frac{g_2(x)}{n^2} = 1 - \frac{\sin^2k_Fx}{(k_Fx)^2},\qquad \lambda=1.
\label{g2lambda1}
\end{eqnarray}
It is interesting to note that the powers characterizing the decay of the
oscillating terms in the pair distribution, Eq.~(\ref{eq:g2_replica}) and in the
one-body density matrix of bosons, Eq.~(\ref{eq:g1_boson_thermo}) are closely
related. Employing the language of electronic systems these oscillating terms can be
referred to as Friedel oscillations. It is a general feature that these oscillations
become more pronounced as one moves to stronger interactions, since they arise due
to the tendency of repelling particles to form a quasi-crystalline
order. Friedel oscillations in the Calogero-Sutherland model have been
discussed in \cite{Nishigaki2003} and similar oscillating behaviour of the same origin has also
been observed in numerical simulations of the strongly interacting limit of
one-dimensional bosons with delta-like interactions \cite{MC_LiebLiniger}.

As interactions are made stronger the amplitude of oscillations become larger and
number of visible oscillations is increased (see Fig.~\ref{Fig:g2}). Although the
quantum fluctuations prohibit the formation of a ``true'' crystal,  this
behaviour of the Calogero-Sutherland model can still be described in terms of the
local crystalline order as was pointed out by Krivnov and Ovchinnikov in
\cite{Krivnov1982}. In the next section we show how the quasi-crystal order
manifests itself in the static structure factor.

\subsection{Static structure factor}

\begin{figure}
\begin{center}
\includegraphics[width=0.5\columnwidth,angle=-90]{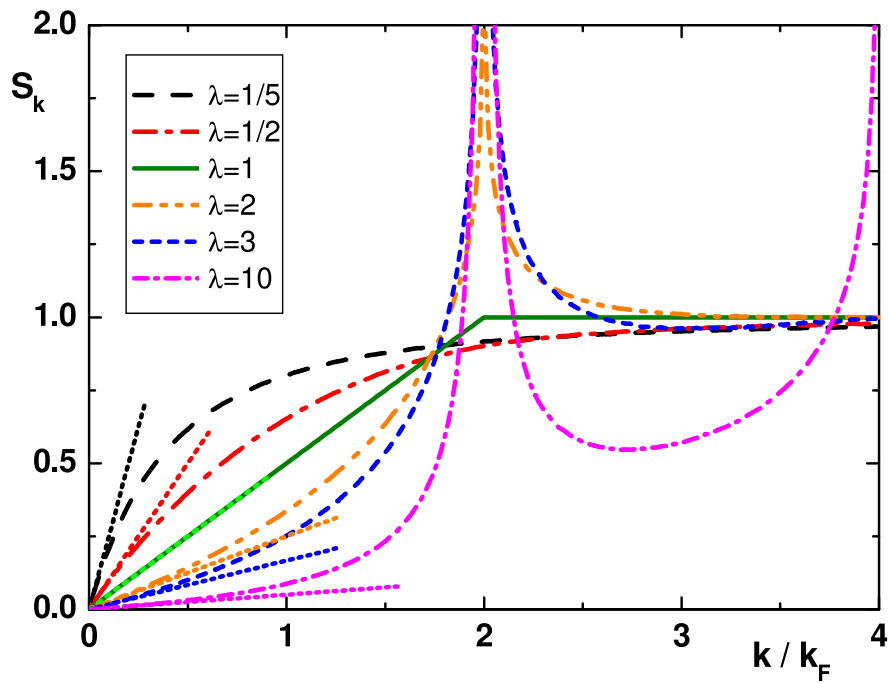}
\caption{(Color online) Static structure factor for different values of the interaction
strength $\lambda$. Dashed lines: low momenta phononic behavior,
Eq.~(\ref{Feynman}). Interaction parameter (in order of decreasing  slope at small $k$):
$\lambda=1/5; 1/2; 1; 2; 3; 10$.}
\label{Fig:Sk}
\end{center}
\end{figure}

Density correlations characterized by the pair distribution function of the previous
subsection are conveniently probed by measuring the static structure factor
(\ref{sk_def}). It is related to the dynamic structure factor $S(k,\omega)$ which
characterizes the scattering cross-section of inelastic reactions where the
scattering probe transfers momentum $\hbar k$ and energy $\hbar\omega$ to the
system. In atomic gases it can be measured directly by the Bragg spectroscopy
technique. By integrating out the $\omega$ dependence one obtains the static
structure factor $S_k$. The numerical results for this quantity in the
Calogero-Sutherland model are summarized in Fig.~\ref{Fig:Sk} for different
interaction strengths.

The behaviour of the static structure factor for small momenta can be described by a
hydrodynamic approach. The Feynman formula \cite{Feynman54} $S_k =
\hbar^2k^2/2m\epsilon_k$ relates the static structure factor to the excitation
spectrum exhausted by one branch $\epsilon_k$. This is the case, when $k$ is small
and the excitations are phonons $\epsilon_k = \hbar k c$ with the speed of
sound~$c$. This leads to linear behavior for small momenta $S_k = \hbar|k|/2mc$.
The speed of sound is related to the compressibility $\chi = mc^2 =
n\partial\mu/\partial n$, where there chemical potential is be found from (\ref{E}) as $\mu
= \partial E/\partial N$. Thus we have
\begin{eqnarray}
S_k = \frac{|k|}{2\lambda k_F}, \qquad k\to 0 ,
\label{Feynman}
\end{eqnarray}
in full agreement with the numerical results as can be seen in Fig.~\ref{Fig:Sk}.
Large momentum excitations behave as free particles $\epsilon_k=\hbar^2k^2/2m$, so
the static structure factor behaves as $S_k \to 1 $ in the limit of large momenta.

In the regime of weak interactions $\lambda\to 0$, the static structure factor is a
smooth function, which goes monotonically from $0$ for $k=0$ to unity for large
values of momenta (see $\lambda=1/5; 1/2$ in Fig.~\ref{Fig:Sk}). This behaviour is
very similar to  that of rarefied weakly interacting gases
\cite{MC_LiebLiniger}.
The critical point  at which a cusp at $k=2k_F$ appears is $\lambda=1$. For
this
critical value of $\lambda$ the static structure factor can be written explicitly
and takes a very simple form:
\begin{eqnarray}
S_k = \left\{
  \begin{array}{ll}
    \frac{|k|}{2k_F}, & |k|<2k_F \\ 1, & |k|\ge 2k_F \\
  \end{array}
  \right.  ,\qquad \lambda=1 .
\label{Sklambda1}
\end{eqnarray}
The linear phononic behaviour continues until $|k|=2k_F$, where the asymptotic
constant value is reached. This special behavior is a result of averaging  the
dynamic form factor $S(k,\omega)$ over all excitation branches (see, for example
\cite{Pitaevskii03}).

For $\lambda>1$ a peak appears in the static structure factor indicating the onset
of the  quasi-crystalline order. In this regime the
correlations between particles on
short distance scale (of order several interparticle separations) become stronger.
Similar physics was observed in numerical simulations of the metastable gas-like
state in a short-range attractive potential (``super-Tonks-Girardeau'' system
\cite{MC_SuperTonks}) or with dipole-dipole interactions \cite{MC_1DDipoles}.

By further increasing the strength of interactions, the peak at
$k=\pm 2k_F$ becomes
higher ($\lambda=2;3;10$ in Fig.~\ref{Fig:Sk}). For extremely strong interactions
several peaks can be observed (see, as an example, $\lambda=10$ in
Fig.~\ref{Fig:Sk}). The absence of coherence observed in the off-diagonal
correlations in this regime allows us to conclude that for extremely large values of
$\lambda$ the Calogero-Sutherland model behaves as a lattice of classical particles,
the one-dimensional Wigner crystal studied in \cite{Krivnov1982}.

\section{Discussion and conclusions}\label{sec:concl}

The present study of the correlation functions of the Calogero-Sutherland model allows
us to make conclusions  about the dominant order, or rather, quasi-order
present in the
system as a function of the interaction parameter $\lambda$. We concentrate on the
case of bosons and identify the following physical regimes at zero temperature.

\begin{figure}
\begin{center}
\includegraphics[width=0.5\columnwidth]{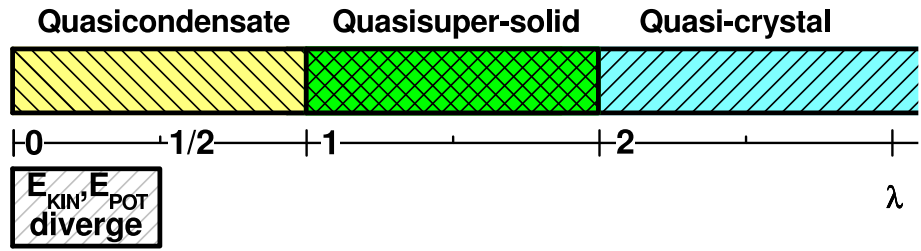}
\caption{(Color online) Phase diagram of the Calogero-Sutherland system.
See text for the explanation.}
\label{Fig:phase diagram}
\end{center}
\end{figure}

For small values of $\lambda$ the system is weakly interacting and stays in a gas-like
(or liquid-like) state. There is a substantial degree of phase coherence in the
system as follows from the slow decay of the off-diagonal correlations. For
$\lambda<2$ the bosonic momentum distribution $n_k$ has an infrared divergence. This
{\it quasi-condensate} regime is  reminiscent of
Bose-Einstein condensation,
where the $k=0$ state is macroscopically occupied. The divergence in the momentum
distribution disappears completely for $\lambda>2$, thus at this interaction value
the Calogero-Sutherland system crosses over from a quasi-condensate to a
non-condensed state for $\lambda=2$ (see Fig.~\ref{Fig:phase diagram}).

By increasing the strength of interactions one finds the appearance of strong
positional ordering of particles ({\it quasi-crystal}) as indicated by the large
amplitude of slowly decaying oscillations in the density-density correlation
function $g_2(x)$. The critical value for this crossover from a liquid to a
quasi-crystal state is estimated as $\lambda=1$, where a singularity in the static
structure $S_k$ appears (see Fig.~\ref{Fig:phase diagram}). For stronger
interactions the particles form a one-dimensional Wigner crystal with a dominant
crystalline order and absence of coherence.

The region $1<\lambda<2$ is very special as the one-body density matrix
exhibits quasi off-diagonal long-range order while there is a
quasi-crystalline order in the static structure factor. We denote this regime,
where those two features are simultaneously present, as a {\it
quasi-super-solid} in analogy with the super-solid state
\cite{AndreevLifshitz1969}. The word ``quasi'' is necessary while talking
about the phase diagrams in a homogeneous one-dimensional system, as there is
no true long-range order and no true phase transitions can take place in such
systems \cite{LandauLifshitzV}. The change of the regimes are actually
crossovers. Indeed, the long-range asymptotics of the one-body density matrix
(\ref{g1B}) include both a slow power-law decay term as well as oscillating
terms for {\it all} values of $\lambda$. The crossover takes place, when one
kind of term becomes dominant over the other terms.

Aside from the question of quasi long-range order related to the behaviour of the
correlation functions for large distances on the scale of the interparticle spacing, we
found interesting phenomena on the scale of the mean distance between particles.
We found intriguing behaviour in the special region $\lambda\le 1/2$ where both the
potential and kinetic energies are divergent, while the total energy and the
compressibility remain finite. The divergence first  occurs for the
critical value $\lambda=1/2$ corresponding to the critical value of
interactions beyond which the fall towards the center takes place \cite{LandauLifshitzIII}.

The presented study of the correlation properties of the Calogero-Sutherland model in
different physical regimes is of a high fundamental interest since this model
provides one of the rare examples of integrable systems for which the correlation
functions can be calculated exactly. Our description holds in physical regimes
ranging from a weakly interacting gas to a strongly correlated crystal-like phase.
We note that in an arbitrary one-dimensional gapless system supporting long-wavelength
phonons, the ground state wave function can be approximately written as (\ref{Phi0})
\cite{Reatto1967}. This general description holds  at distances
where the
hydrodynamic approach is applicable. So, the study of the Calogero-Sutherland model is
important for understanding the long-wavelength properties of systems with linear
low-momenta excitation spectra (also known as Luttinger liquids).

Apart from its theoretical importance, the Calogero-Sutherland model is relevant
for several realistic physical systems. The classic example
is provided by the compressible borders of Fractional Quantum Hall droplets \cite{Stone1994}.
The experiments on vicinal crystal surfaces \cite{Einstein2003} provide yet another
physical realization of CSM. Recently, the particular case $\lambda=1$ of bosonic
CSM has been realized in the series of experiments \cite{TG:experiments} with tightly
confined cold atomic gases.

It is important to stress that regardless of the absence of the true long-range
order, the dominant quasi-order can reveal itself in mesoscopic confined systems due
to their finite size. The study of physical regimes in mesoscopic systems
confined in harmonic potentials are certainly of interest  and will be
considered in our future studies. Among other open questions one can name the
time-dependent correlations as well as  correlation functions of
the Calogero-Sutherland model at finite temperature.

\section{Acknowledgments}

D.M.G. would like to thank the University of Trento and the Institute of
Spectroscopy for
their hospitality during the work on this project. G.E.A. is grateful to LPTMS for
hospitality. The work was supported by grants from the Minist\`ere de la
Recherche (grant ACI Nanoscience 201), Agence Nationale de la Recherche, IFRAF
Ministero dell'Istruzione, dell'Universit\`a e della Ricerca (MIUR) and RFBR.
We thank B.~Jackson for reading the manuscript. LPTMS is the mixed research
unit No. 8626 of CNRS and Universit\'e Paris
Sud.

\appendix

 \section{Analytical continuation of $A_n$}
 \label{app:analAn}

We show how to perform an analytic continuation of the constant
\begin{eqnarray}
  \label{eq:am}
  A_m^2(\lambda) = \frac{\Gamma(m+1)}{\Gamma^2\left(m/2+1\right)}
  \prod_{c=1}^m\Gamma\left(\frac{m+c}{\lambda} \right) \prod_{a=1}^{m/2}
  \frac{\Gamma\left(1+\frac{a}{\lambda}\right)}
  {\Gamma\left(1+\frac{m/2+a}{\lambda}\right)\Gamma^2\left(\frac{m/2+a}{\lambda}\right)}
\end{eqnarray}
to $m=\lambda$. The first factor in the right hand side of this expression can
be continued straightforwardly.  Consider the logarithm of $A^2_m$
 \begin{equation}
   \label{eq:logAm}
   \ln \left(A^2_m \frac{\Gamma^2(m/2+1)}{\Gamma(m+1)}\right)=
\sum_{c=1}^{m}\ln \Gamma \left(\frac{m+c}{\lambda}\right)+
\sum_{c=a}^{m/2}\left(\ln \Gamma \left(1+\frac{a}{\lambda}\right)- \ln \Gamma
\left(1+\frac{m/2+a}{\lambda}\right)-2 \ln \Gamma
\left(\frac{m/2+a}{\lambda}\right)\right)
 \end{equation}

We use the following integral representation \cite{GradshteynRyzhik} for the
logarithm of Euler's gamma function
 \begin{equation}
   \label{eq:log_gamma}
   \ln \Gamma (z) = \int_0^\infty \frac{dt}{t}
   \left(\frac{e^{-zt}-e^{-t}}{1-e^{-t}}+(z-1) e^{-t}\right)
 \end{equation}
to represent each term in the right hand side of (\ref{eq:logAm}).  Summing
finite geometric and arithmetic series under the integral we get
 \begin{equation}
   \label{eq:logAm_int}
   \ln \left(A^2_m \frac{\Gamma^2(m/2+1)}{\Gamma(m+1)}\right)= \int_0^\infty
\frac{dt}{t} e^{-t} \left(\frac{m^2}{2\lambda}+
\frac{\left(1-e^{-\frac{mt}{2\lambda}}\right)^2 - e^t
\left(e^{-\frac{2mt}{\lambda}}-3e^{-\frac{mt}{\lambda}}+2e^{-\frac{mt}{2\lambda}}\right)}
{\left(1-e^{-t}\right)\left(e^{t/\lambda}-1\right)}\right),
 \end{equation}
which under replacement $m=\lambda$ yields the result (\ref{eq:alambda}).

The calculation in the fermionic case is similar. We have
\begin{eqnarray}
  \label{eq:cm}
  C^2_m (\lambda)
  =\frac{\Gamma(m+1)}{\lambda}\frac{\Gamma^2\left(\frac{m+1}{2\lambda}\right)}
  {\Gamma^2\left(\frac{m+1}{2}\right) }\times \frac{\prod_{c=1}^m
  \Gamma\left(\frac{m+c}{2}\right) \prod_{a=1}^{\frac{m-1}{2}}
  \Gamma\left(1+\frac{a}{\lambda}\right)} {\prod_{b=1}^{\frac{m+1}{2}}
  \Gamma\left(1+\frac{m-1+2b}{2\lambda}\right)
  \Gamma^2\left(\frac{m-1+2b}{2\lambda}\right)}.
\end{eqnarray}
Taking the  logarithm of this expression, using the integral
representation (\ref{eq:log_gamma}) and putting $m=\lambda$ yields the result (\ref{eq:C}).

\end{document}